\documentclass[3p]{elsarticle}
\usepackage{graphicx}
\usepackage{bm}
\usepackage{mathptmx}
\usepackage{epstopdf}
\usepackage{booktabs}
\usepackage{multirow}
\usepackage{hhline}
\usepackage{dcolumn}
\usepackage{subfigure}
\usepackage{amsmath}
\usepackage{float}
\usepackage[usenames,dvipsnames]{color}

\begin{document}
\begin{frontmatter}
\title{NH$_3$ adsorption on PtM (Fe,~Co,~Ni) surfaces: cooperating effects of charge transfer, magnetic ordering and lattice strain}
\author{Satadeep Bhattacharjee$^{1}$, S. J. Yoo$^{2}$, Umesh V. Waghmare$^{3}$ and S. C. Lee$^{1,4}$}
 \address{$^{1}$Indo-Korea Science and Technology Center (IKST), Bangalore, India \\
    $^{2}$Fuel Cell Research Center, Korea Institute of Science, Korea\\
    $^{3}$Jawaharlal Nehru Centre for Advanced Scientific Research(JNCASR), Bangalore, India \\
   
    $^{4}$Electronic Materials Research Center, Korea Institute of Science $\&$ Tech, Korea}

\begin{abstract}Adsorption of a molecule or group with an atom which is less electronegative than oxygen (O) and directly interacting with the surface is very relevant to development of PtM~(M=3d-transition metal) catalysts with high activity. Here, we present theoretical analysis of the adsorption of NH$_3$ molecule (N being less electronegative than O) on (111) surfaces of PtM(Fe,Co,Ni) alloys using the first principles density functional approach. We find that, while NH$_3$-Pt interaction is stronger than that of NH$_3$ with the elemental M-surfaces, it is weaker than the strength of interaction of NH$_3$ with M-site on the surface of PtM alloy. 
\noindent
\end{abstract}
\end{frontmatter}
Bimetallic surfaces of Pt alloyed with first row transition metals (M) such as Fe,~Co,~Ni etc. are interesting and promising 
for their potential applications as  catalytic cathodes in Proton Exchange Membrane Fuel Cell (PEMFC). These alloys not only ensure the cost effectiveness but also reduce the over-potential and increase the oxidation reduction reaction (ORR) activity\cite{imp1,imp2,imp3}. In a cathode which is made up of pure Pt, the accumulation of oxygen species, such as  O, OH etc, decrease the availability of active sites, as well as  increase the activation barrier for the ORR.  In recent study it was shown  that carbon supported PtM alloys exhibit improved catalytic activity by preventing the oxidation and the dissolution of the Pt in the aqueous medium\cite{imp2}. 
However, yet the catalytic activity is much weaker than what is expected theoretically. This is because, the oxophilic nature of M atoms results in the formation of surface M-oxides leading to degradation of catalytic activity of these bi-metals. 
To prevent such effects, it was proposed that selective coverage of M-sites with ligands containing lesser electro-negative elements such as nitrogen (N) would reduce the oxidation of M species and thereby help more Pt sites to remain active \cite{Jung}. The enhancement of catalytic activity along this route involves two key steps: (1) proper selection of M which can provide Pt-sites suitable environment to retain their ORR activity and (2) tuning the electronic structure of M-sites in order to prevent formation of M-oxides.
\par
The \textit{first step} is emphasized as follows: the electro negativity difference between M and Pt causes the charge transfer from M to Pt site, resulting a down-shift of the Pt-d states via filling of the Pt d-band, thereby reducing the accumulation of oxygen species over Pt sites. To achieve the \textit{second step}, one requires preferential adsorption of N-containing ligand on M-sites, i,e  M should be \textit{more N-philic} than Pt, in a PtM alloy condition.
\par
The electronic, chemical and structural properties of the transition metal (M) are therefore very important to achieving this two mechanisms. The predictability of the modification of electronic and chemical properties of M in an \textit{alloy environment} with respect to its elemental state is a challenging task. It is therefore necessary  to find the descriptor(s) which properly indicate the changes in electronic, magnetic as well as chemical behaviour  
of  these transition metals in the alloyed form. The same descriptor(s) can also be used in prediction of  the catalytic activity of the PtM catalyst. Change in catalytic activity of transition metal surfaces  or more precisely the adsorption energy of small molecules are described in terms of so called d-band center model of Hammer and N{\o}rskov\cite{d-band1,d-band2,d-band3}, which predicts a linear correlation between the shift of the d-band center of one metal with respect to the other with the energy of adsorption. Predictions of chemisorption properties in terms of the center of the d-band which is an energy point averaged over spin-up and spin-down d-band centers. therefore the effects of the spin polarization are neglected. In PtM alloys, the d band filling of the metals (M) are different from their parent state. The lattice mismatch between the M-Pt results a strain in the system, which can also influence the catalytic activity. 
\par
To understand the possibility of selective M-coverage with  N-containing ligand, we have studied the adsorption of NH$_3$  on PtM (Fe,~Co,~Ni) surfaces. Our calculations reveals a reversibility of adsorption behaviour of NH$_3$ molecule on M sites of PtM in comparison to the pure M(Fe,Co,Ni) surfaces. 
NH$_3$ has larger adsorption energies on M-site of PtM surface than of an M-site of elemental surfaces. We relate the  such changes in adsorption energies to the physical mechanisms such as charge transfer, magnetic moment and strain effects. Although the effects of charge transfer and strain can be related to the shift of the d-band center, the effect spin polarization is not included in the d-band model. Most prominently, we have quantified the effects of these mechanisms from our first-principles calculations. For designing highly efficient alloy catalysts such as PtM, we suggest that one should gather insights from a multicomponent descriptor rather than a scaler descriptor such as d-band center.
\par
Our first-principles calculations were within the framework of density functional theory (DFT) with 
Perdew-Burke Ernzerhof functional of exchange correlation energy derived within a 
generalized gradient approximation\cite{pbe} and projector augmented wave (PAW) method 
as implemented in Vienna \textit{ab initio} simulation package (VASP)\cite{vasp}. 
The PAW potentials include the following valence electrons: 3d$^8$4s$^1$ for Co, 5d$^9$5s$^1$ for Pt, 
2s$^2$2p$^3$ for N, and 1s$^1$ for H. Surfaces of Pt, M and PtM were simulated using periodic supercells
containing slabs of their $4\times 4$ in-plane unit cells and thickness of four atomic planes, thus consisting of
64 atoms. The top two atomic planes were relaxed while the bottom two atomic planes were kept fixed 
in their bulk structure. Wavefunctions of valence electrons were expanded in a plane wave basis set
truncated with an energy cut-off of 450 eV. The integrations over the Brillouin zone were sampled on
uniform grid of $3\times 3\times 1$ k-points using Methfessel-Paxton smearing with a width of 0.1 eV.  Ionic relaxation are performed such that the force on each ion is smaller than 0.02 eV/\AA. The dipole correction was applied along the direction 
perpendicular to the metal surfaces to correct the electric fields arising from the structural
asymmetry and periodicity. The adsorption energies were determined using the relation,
\begin{equation}
E_{ad}=E_{S+A}-(E_S+E_A),
\end{equation}
where $E_{S+A}$ is the energy of the surface plus adsorbate while $E_S$ and $E_A$ are  energies 
of the surface and adsorbate respectively. Magnetic interactions among  M-M and M-Pt are considered to be ferromagnetic.    
\par
For each PtM alloy considered here, we first obtained its optimized structure for (1:1) composition. We optimized its structure with tetragonal 
symmetry with P4/mmm space group. In the optimized structure, (see Table. I),  $\frac{c}{a}\simeq \sqrt{2}$, and M-atom occupies the 1a (0,0,0) site and Pt occupies 1d (0.5,0.5,0.5) site. It is known that the L1$_0$ phase become stable for FePt below 1300 $^\circ$C, for CoPt below 825 $^\circ$C and for NiPt below 1300 $^\circ$C \cite{phase}. The (111) surface was constructed from this relaxed bulk structure. In the Table II, we report the adsorption energies of NH$_3$ molecule 
on elemental metal surfaces as well as on MPt surfaces for the \textit{atop} positions, which is the most preferred adsorption site for NH$_3$ on the transition metal surfaces \cite{PCCP,atop,TM2}. For the elemental metal surfaces, NH$_3$ has highest adsorption energy on Pt (111) surface followed by Ni (111), Fe (110) and Co (0001). However, on MPt surfaces NH$_3$ is found to have stronger binding at M sites than Pt. The strength of NH$_3$-Pt binding on MPt surfaces decrease as we move from Ni to Fe. Such reversal of adsorption behaviour is interesting and also give us scope for tuning the activity of PtM alloy through manipulating the electronic structure of M-sites using suitable ligands. The question is: how do we understand such reversal of chemisorption?
\par
From the average magnetic moments at the M-sites in PtM surfaces and those in the pure M-surfaces (shown in bracket in the Table I),  we see that the magnetic moments on M-sites are larger in PtM than in pure M. This is due to charge transfer from the minority d-band of M to d-band of Pt as is evident in spin-resolved Bader charges\cite{Bader} reported in Table III. It is seen that about 0.82 electrons transfer from Fe-d bands to Pt d-bands, of which is 0.45 electrons transfer to the majority spin channel and 0.37 electrons are transfer to the down spin channel of the Pt-d bands. Similarly, 0.48 electrons  transfer from Co site to the Pt site, of which 0.41 electrons  transfer to the majority spin channel and only 0.07 electrons of Pt-d band. In the case of NiPt, 0.34 electrons transfer from Ni to Pt, out of which 0.31 electrons are transferred to to the majority Pt-d band and only 0.03 electrons transfer to Pt minority d-band. This charge transfer  also induces a small magnetic moment on the Pt atom, the magnitude of the magnetic moment being 0.5, 0.45 and 0.34 $\mu_B$ respectively for FePt,CoPt and NiPt. 
\par
From the optimized lattice constants (Table. I) we expect strain effects, as reflected in  the magnitude of the surface vectors for the 
pure metal (M) as well as PtM surfaces (Table. IV), it is clear that the M-M distances are longer for PtM than that in pure M surface (by
about 8\% for Fe, 6\% for Co and 7\% for Ni). According to the d-band model of heterogeneous catalysts, for the late transition metals such as Fe,Co and Ni the increase in inter atomic distance for M lattice leads to narrowing of M d-band width which  
will cause the increase in the strength of chemisorption due to an upward shift of the d-band center to preserve the charge 
conservation\cite{gross}.
\par 
From the projected density of states (DOS), it is clear that the majority spin d-bands are completely full in FePt and CoPt and determine the M-magnetic moment while magnetic moment is determined by both majority and minority spin electrons in NiPt.  
\par
In pure metallic surfaces there is no inter-site charge transfer since there is no difference in the electronegativity among these
sites. However for the PtM systems there is charge transfer effect arising from the difference in the electro negativity of M 
and Pt atoms. On the Pauling scale, the electronegativity difference between Pt and M (Fe,~Co,~Pt) are  0.45,0.40,0.37 respectively. The magnetic moments on M-sites in PtM surfaces are in general larger than what is there in M-sites in a pure M-surface. Also there are strain effects in due to the lattice mismatch between Pt and M in PtM. The difference in the adsorption energy of NH$_3$ molecule from M to the PtM surfaces arise due to a combined effects of the above mentioned phenomena. To uncover the  individual contributions of  charge transfer, magnetism and strain we apply the following methodology.
\par
The adsorption energy of the NH$_3$ molecule on the M surface is given by,
\begin{equation}
E_{ad}^M=E_{M+A}-E_M-E_A
\end{equation}
while that on a PtM surface is written as,
\begin{equation}
E_{ad}^{PtM}=E_{PtM+A}-E_{PtM}-E_A
\end{equation}
The change of the adsorption energy of NH$_3$ molecule from the M to PtM surface can be written as\cite{PCCP}
\begin{align}
\Delta E_{ad}&=E_{ad}^{PtM}-E_{ad}^M=(E_{PtM+A}-E_{M+A})- (E_{PtM}-E_M)\nonumber \\
             &=\Delta E_q+\Delta E_{\bf m}+\Delta E_{\bf \varepsilon},
\end{align}
where $\Delta E_q$, $\Delta E_{\bf m}$ and $\Delta E_{\bf \varepsilon}$ are respectively the contributions to the 
changes in adsorption energy $\Delta E_{ad}$ coming from the charge transfer, magnetic moments and strain. To calculate 
$\Delta E_q$, $\Delta E_{\bf m}$ and $\Delta E_{\bf \varepsilon}$ we have used following method: We assume that one can simulate the effects of Pt-coordination on M, by considering the pure M surface with an enhanced surface area (to capture the effect of strain), a larger magnetic moment and a less number of electrons per M atom. We express such assumptions in a mathematical form given by,
\begin{equation}
\begin{split}
\Delta E_{\bf \varepsilon}=E_{ad}^M(S_{PtM})-E_{ad}^M(S_M)\\
\Delta E_{\bf  m}=E_{ad}^M({\bf m}_{PtM})-E_{ad}^M({\bf m}_M)\\
\Delta E_{q}=E_{ad}^M(q_{PtM})-E_{ad}^M(q_M)
\end{split}
\end{equation}
$\Delta E_{\bf \varepsilon}$ involves calculation of adsorption energy of NH$_3$ on the M-surface at its equilibrium surface area $S_M$ and on the M-surface with enhanced surface area $S_{PtM}$ (which is the surface area of PtM). 
To obtain $\Delta E_{\bf  m}$, we perform calculation of adsorption energies for the M-surface surface with an enhanced magnetic moment, by con-straining the overall magnetic moment of the system such that the magnetic moment per M ion is about ${\bf m}_{PtM}$ (which is the magnetic moment per M ion in a PtM surface). Since there is no change in the electronic charge inside the M sphere (only redistribution of majority and minority spin  electrons), no charge transfer effect is involved. Also since the structure is unchanged, the strain effects
is absent too. The last term of the Eq.5 ($\Delta E_{q}$) is obtained by subtracting the magnetic and lattice contributions from the total $\Delta E_{ad}$ (from the Eq.(4)).
\par
Using the methodologies above, we have split the $\Delta E_{ad}$ into its lattice, magnetic and charge transfer contributions (see results tabulated in Table. IV). It is seen that for FePt and CoPt the increase of NH$_3$ adsorption energy w.r.t the Fe and Co surface can be related mostly to the strain and the charge transfer effects, magnetism plays weaker role here. For NiPt, due to relative smaller difference of electro negativity between Ni and Pt the charge transfer effect is small. The change in adsorption energy is mainly dominated by strain effect. Also, among all three, FePt is expected to be best in terms of activity since the selective M-coverage with NH$_3$ is easiest in this case. 
\par
In the Fig.2, we show the DOS of NH$_3$ in the gas phase (2a), the Co-d projected DOS and NH$_3$ projected DOS for NH$_3$ adsorbed Co (0001) surface (2b) and d-projected DOS for the Co (0001) surface (2c). From 2(a) and 2(c) it is evident that the lowest unoccupied levels (4a1,2e levels) of NH$_3$ are well above the d-states and are weakly involved in the chemisorption. It is the non-bonding lone pair (3a1 level) which describes the M-NH$_3$ bonding. On the metal surface the lone pair is highly stabilized (by about 4 eV), while the three NH$_3$ bonds (doubly degenerated 1e level and 2a1 level) are also stabilized slightly.
\par
Since NH$_3$ molecule interacts with the surface through the lone-pair electrons, it is easy to understand why NH$_3$ prefers to M-sites in PtM surface. The lone pair 
electrons prefer to interact more strongly to the electron deficient M-atoms  than Pt atoms which are more negatively charged due to electron transfer from the M-atoms. This is also demonstrated in the Fig.3, where we show that DOS of the lone pair on CoPt surface. The lone pair is mainly composed of N-P$_z$ orbital with little contribution from N-s orbital\cite{TM3}. From the figure it can be understood that, when the  NH$_3$ molecule is adsorbed on one of the Co, it donates lone pair electrons to that Co and therefore lone-pair-DOS moves \textit{upwards} in energy w.r.t the case when it is adsorbed on Pt-site. NH$_3$  molecule is therefore less nucleophilic on negatively charged Pt sites.
\par
In conclusion, we have studied the cooperative roles of charge transfer, magnetism and strain in adsorption of NH$_3$ on FePt, CoPt and NiPt surfaces. Since NH$_3$ has a larger adsorption energy on M-sites compared to Pt sites, it is possible to achieve the selective NH$_3$ coverage of M sites on PtM surfaces. Such selective coverage is easiest in the case of FePt due to the largest M spin moments, strongest charge transfer and associated strain effects. Among the three mechanisms, the effects of charge transfer and strain are the largest, while the contribution of magnetism is relatively modest. We therefore predict FePt to be the most effective among the three. In principle, one can propose a three component descriptor ${\bf D}={\bf D}(\Delta q,\Delta {\bf m},\Delta {\bf \varepsilon})$ , where three components of the descriptor measure the change in charge, magnetic moment and strain effects in PtM surfaces compared to that of elemental M-surface.
\par
This work was supported by NRF grant funded by MSIP, Korea(No. 2009-0082471 and No. 2014R1A2A2A04003865) and the Convergence Agenda Program (CAP) of the Korea Research Council of Fundamental Science and Technology (KRCF). UVW acknowledges funding from the Indo-Korea Institute of Science and Technology (IKST).        

\newpage
%
\begin{table}
\begin{tabular}{|c|c|c|c|}
\hline 
{\bf PtM} &  \multicolumn{2}{|c|}{\bf Lattice constants} & {\bf Magnetic moments}\\ 
\hline
    &   a ($\AA$) & c ($\AA$) & on M-site ($\mu_B$)\\
\hline
FePt& 2.70 & 3.82 & 3.01 (2.58)\\ 
\hline 
CoPt & 2.65 & 3.74 & 1.9 (1.7) \\ 
\hline 
NiPt & 2.67 & 3.78 & 0.8 (0.7) \\  
\hline 
\end{tabular}
\caption{The lattice constants and magnetic moments on the M-site of M(Fe,Co,Ni)Pt in their L1$_0$. In the bracket, we also
show the values of the magnetic moments for pure M-surfaces.} 
\end{table}
\begin{table}
\begin{tabular}{|c|c|c|}
\hline 
{\bf Surface} & {\bf Adsorption Site} & {\bf Adsorption energies (eV)} \\ 
\hline 
\hline
Fe (110) & Fe & -0.72\\ 
\hline 
Co (0001) & Co & -0.69 \\ 
\hline 
Ni (111) & Ni & -0.80\\ 
\hline
Pt (111) & Pt & -0.95\\
\hline
\hline 
FePt & Fe & -0.85 \\ 
     &  Pt &  -0.53 \\ 
\hline 
CoPt & Co & -0.80\\ 
     & Pt  &  -0.66 \\ 
\hline 
NiPt & Ni & -0.87\\ 
     & Pt &  -0.74\\ 
\hline 
\end{tabular} 
\caption{Calculated adsorption energies on  'on top' sites for different transition metal surfaces.}
\end{table}
\newpage
\begin{table} 
\begin{tabular}{|c|c|c|c|c|}
\hline 
{\bf Surface} & {\bf Site} & {\bf n$_\uparrow$} & {\bf n$_\downarrow$} & {\bf (n$_\uparrow$+n$_\downarrow$)} \\ 
\hline
\hline 
Pt & Pt & 5.00& 5.00 & 10.00 \\ 
\hline 
Fe (011) & Fe & 5.30 & 2.71 & 8.01 \\ 
\hline 
Co (0001) & Co & 5.35 & 3.65 & 9.00\\ 
\hline 
Ni (111) & Ni & 5.32 & 4.68 & 10.00 \\ 
\hline 
\hline
FePt & Pt & 5.45 & 5.37 & 10.82 \\ 
 & Fe & 5.32 & 1.86 & 7.18 \\ 
\hline 
CoPt & Pt & 5.41 & 5.07 & 10.48 \\ 
 
& Co & 5.22 & 3.30 & 8.52 \\ 
\hline 
NiPt & Pt & 5.31 & 5.03 & 10.34\\ 

 & Ni & 5.22 & 4.43 & 9.65 \\ 
\hline 
\end{tabular} 
\caption{Spin resolved Bader charge per atom for different atoms on different surfaces.}
\end{table}
\begin{table}
\begin{tabular}{|c|c|c|}
\hline 
{\bf Metal} & \multicolumn{2}{c|}{\bf Magnitude of the surface vector in \AA} \\ 
 & {\bf M} & {\bf PtM} \\ 
\hline 
Fe & (2.87,2.02)& (2.70,2.70) \\ 
\hline 
Co & (2.50,2.50) & (2.65,2.65)\\ 
\hline 
Ni & (2.48,2.48) & (2.67,2.67)\\
\hline
\end{tabular} 
\end{table}

\begin{table}
\begin{tabular}{|c|c|c|c|c|}
\hline 
{\bf PtM} &{\bf $\Delta E_{ad}$} & {\bf $\Delta E_q$} & {\bf $\Delta E_{\bf m}$} & {\bf $\Delta E_{\epsilon}$} \\ 
    &  (meV)  & (meV) & (meV) & (meV) \\
\hline
FePt & 130 & 50 & 30 & 50 \\ 
\hline 
CoPt & 110 & 40 & 20 & 50 \\ 
\hline 
NiPt & 70 & 20 & 10 & 40 \\ 
\hline 
\end{tabular} 
\caption{The role of charge transfer, magnetism and strain effects in the reversal behaviour of adsorption.}
\end{table}
\clearpage
\newpage 
\begin{figure}
\includegraphics[width=130mm,height=120mm]{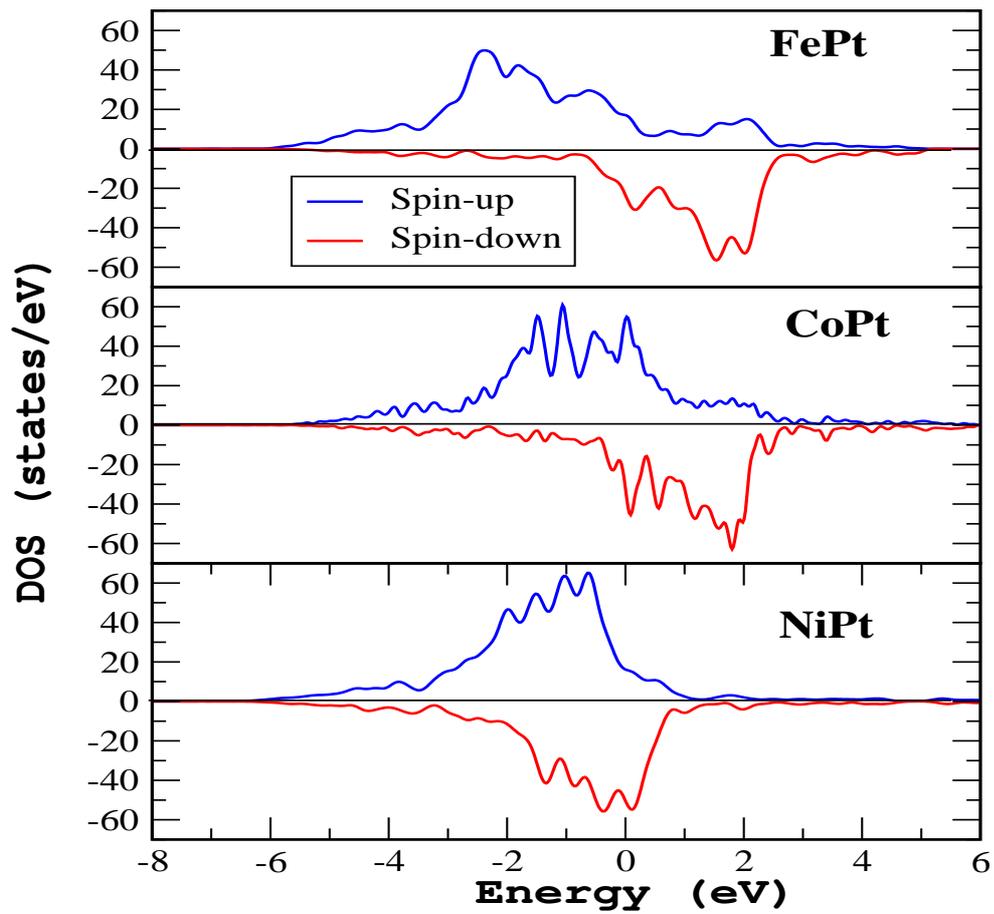}
\caption{(Color online) The density of states projected on M (Fe,Co,Ni) d-states of FePt,CoPt and NiPt}
\label{Fig.1}
\end{figure}
\newpage 
\begin{figure}
\includegraphics[width=160mm,height=140mm]{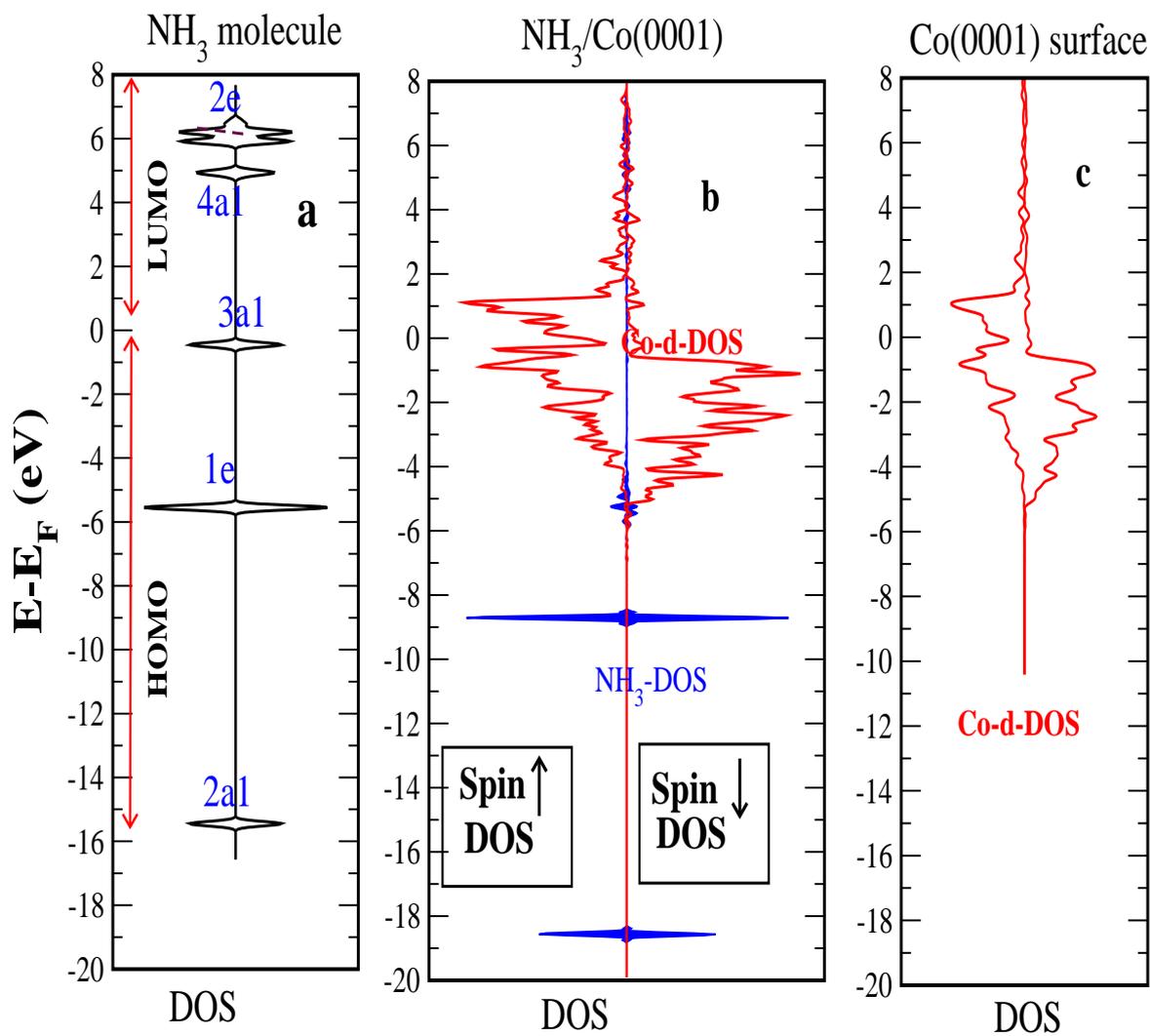}
\caption{(Color online) (a)~The density of states of NH$_3$ molecule in gas phase (b)~the Co-d projected DOS and NH3 projected DOS for NH3 adsorbed Co (0001) surface and (c)~d-projected DOS for the Co (0001)}
\label{Fig.2}
\end{figure}
\newpage 
\begin{figure}
\includegraphics[width=160mm,height=140mm]{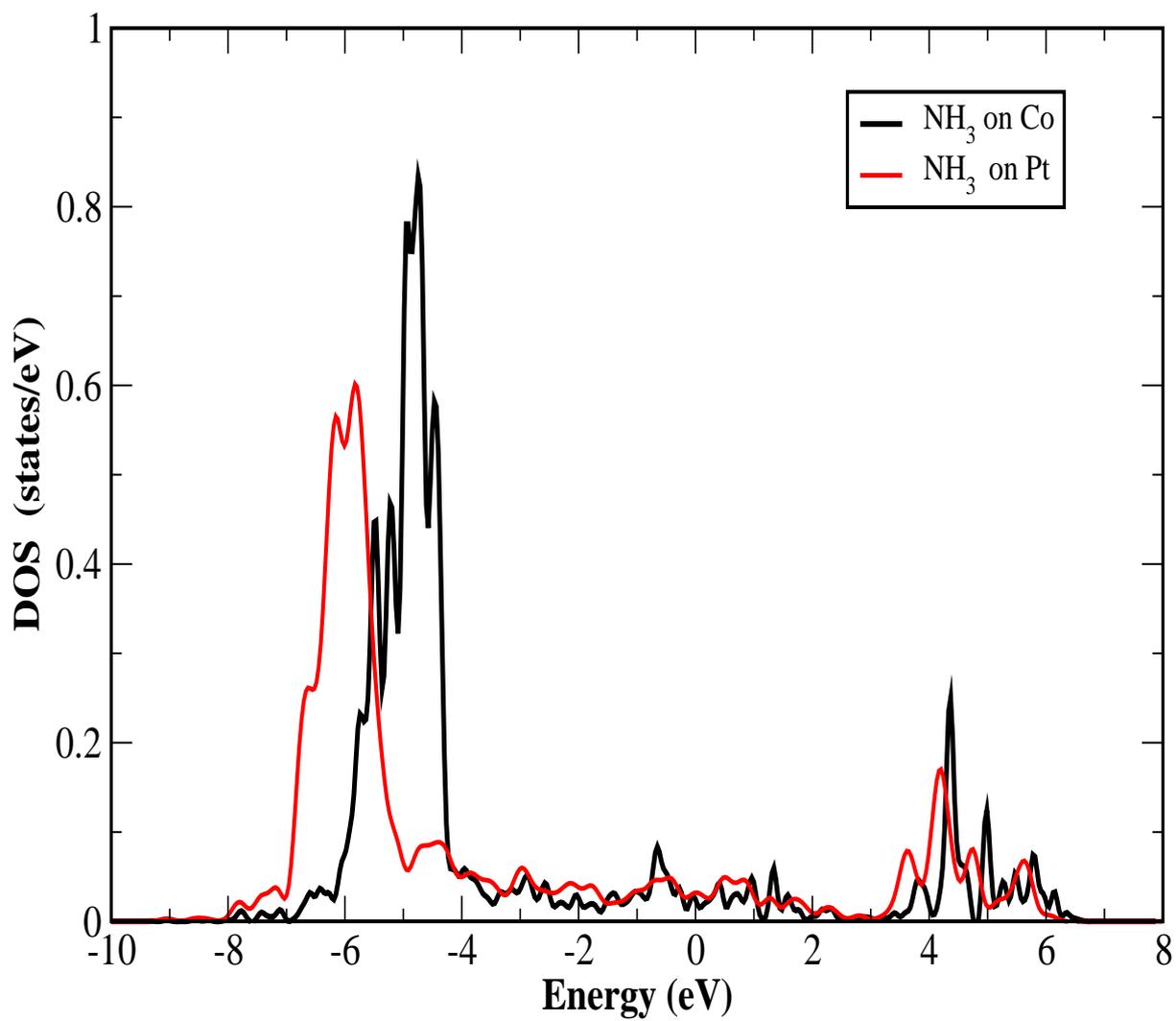}
\caption{(Color online) The density of states projected to the lone pair of NH$_3$ molecule adsorbed on CoPt surface.}
\label{Fig.3}
\end{figure}
\newpage 
\begin{figure}
\includegraphics[width=160mm,height=140mm]{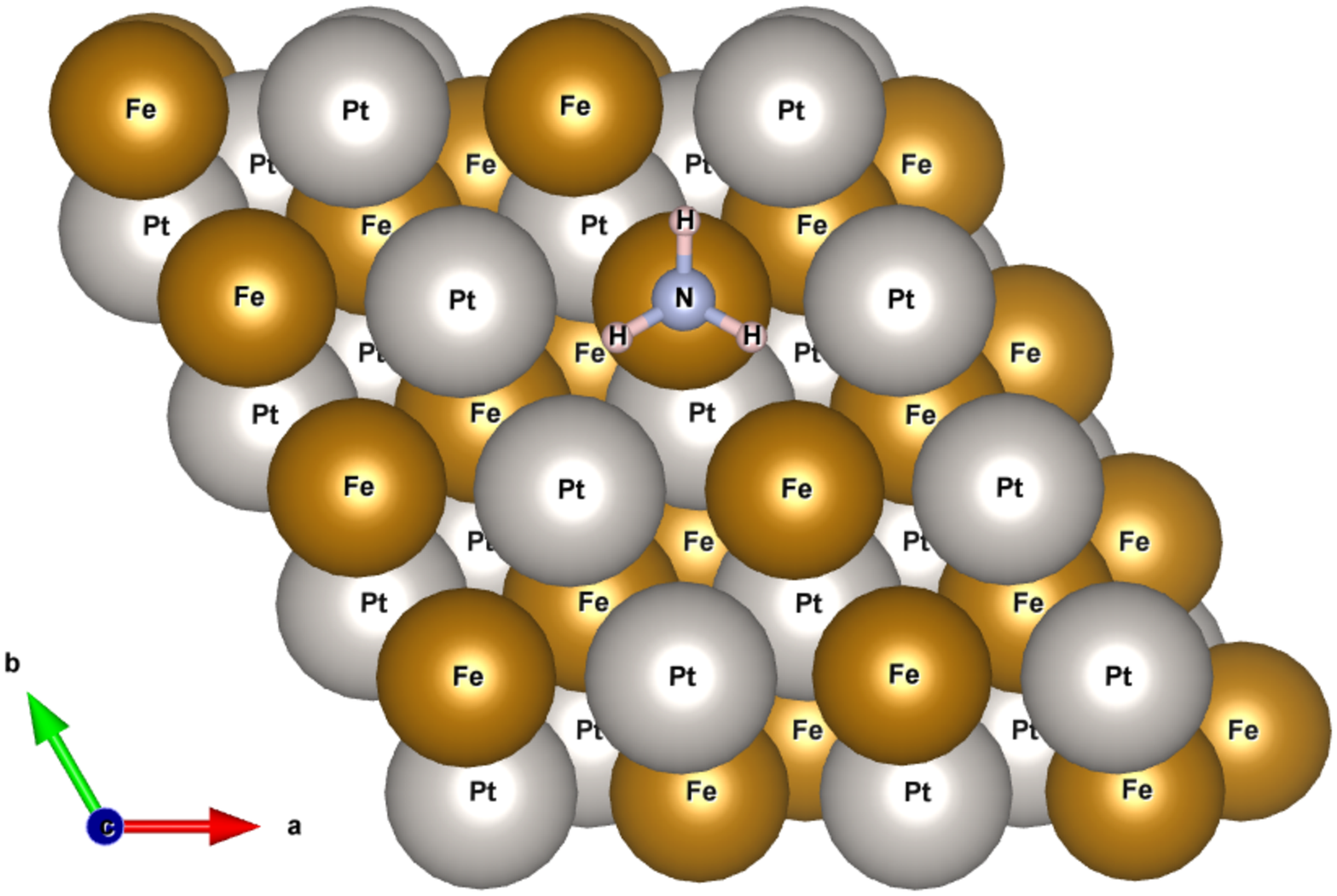}
\caption{(Color online) NH$_3$ adsorbed on FePt surface (viewed along c-axis) as a representative of the adsorption geometry.}
\label{Fig.3}
\end{figure}


\newpage
\begin{thebibliography}{99}
\bibitem{imp1}E. Antolini, J. R.C. Salgado , R. M. da Silva, E. R. Gonzalez, Materials Chemistry and Physics, {\bf 101}, 395-403 (2007)
\bibitem{imp2}K. Hyun, J. H. Lee, C. W. Yoon and Y. Kwon, Int. J. Electrochem. Sci, {\bf 8}, 11752-11767 (2013) 
\bibitem{imp3}D. He, S. Mu and S. Pan, Carbon, {\bf 49}, 82-88 (2011)
\bibitem{Jung} N.Jung,\textit{et. al},(Accepted in NPG Asia materials, 2015)
\bibitem{d-band1}B. Hammer and J.K. N{\o}rskov, Surf. Sci. {\bf 343}, 211-220 (1995)
\bibitem{d-band2}B. Hammer and J.K. N{\o}rskov, Nature, {\bf 376}, 238-240 (1995)
\bibitem{d-band3}B. Hammer and J.K. N{\o}rskov, Advanes in catalysis, {\bf 45}, 71-129 (2000)
\bibitem{phase} TB Massalski , H. Okamoto, \textit{Binary alloy phase diagrams [CDROM]. Materials Park, OH: ASM International} 1996
\bibitem{PCCP}S.Bhattacharjee, K. Gupta, N. Jung, S. J. Yoo, U.V. Waghmare and S. C. Lee,~Physical Chemistry Chemical Physics, {\bf 17}, 9335-9340 (2015)
\bibitem{atop} G. Novell-Leruth,A. Valcarcel, J. Perez-Ramırez, and J. M. Ricart, J. Phys. Chem. C, {\bf 111}, 860-868 2007
\bibitem{TM2}Theoretical Aspects of Transition Metal Catalysis, Topics in Organometallic Chemistry, vol 12 , \textit{Springer}, 2005
\bibitem{Bader} W. Tang, E. Sanville,  G. Henkelman, J. Phys.: Condens. Matter,  {\bf 21}, 084204-084204 2009,
\bibitem{gross}S. Schnur and A. Gro{$\beta$}, Phys. Rev. B, {\bf 81},  033402(1)-033402(4) (2010)
\bibitem{TM3}H. Cheng, D. B. Reiser, P. M. Mathias, K.  Baumert, and Sheldon W. Dean, Jr., J. Phys. Chem., {\bf 99}, 3715-3722 1995
\bibitem{pbe} J. P Perdew, K. Burke, M. Ernzerhof, ~Physical Review Letters, {\bf 77}, 3865-3868 (1996)
\bibitem{vasp}G. Kresse, J. Furthm\"uller,~Physical Review B, {\bf 54}, 11169-11186 (1996)
\end{thebibliography}
\end{document}